\documentclass[aps,physrev,reprint,superscriptaddress]{revtex4-2}
\usepackage[colorlinks=true,
            citecolor=blue,
            linkcolor=blue,
            urlcolor=blue]{hyperref}
\usepackage{graphicx}

\usepackage{amsmath}

\usepackage{mathrsfs}

\usepackage{float}

\usepackage{placeins}

\usepackage{amssymb}

\usepackage{cleveref}

\begin{document}

\title{Interpretation of chirality imprinted on photoelectron momentum spectra by three-dimensional light fields}

\author{Junxin Yuan}
\affiliation{Beijing National Laboratory for Condensed Matter Physics, Institute of Physics, Chinese Academy of Sciences, Beijing 100190, People's Republic of China}
\affiliation{University of Chinese Academy of Sciences, Beijing 100049, People's Republic of China}

\author{Yuhong Li}
\affiliation{School of Electrical Engineering, Lanzhou Institute of Technology, Gansu 730050, People's Republic of China}

\author{Fei Li}
\affiliation{Department of Physics, College of Science, Shantou University, Shantou, Guangdong 515063, People's Republic of China}

\author{Jinkai Zhang}
\affiliation{Beijing National Laboratory for Condensed Matter Physics, Institute of Physics, Chinese Academy of Sciences, Beijing 100190, People's Republic of China}
\affiliation{Institute of Theoretical Physics, State Key Laboratory of Quantum Optics and Quantum Optics Devices, Taiyuan 030006, People's Republic of China}

\author{Shiyang Zhong}
\affiliation{Beijing National Laboratory for Condensed Matter Physics, Institute of Physics, Chinese Academy of Sciences, Beijing 100190, People's Republic of China}

\author{Hao Teng}
\affiliation{Beijing National Laboratory for Condensed Matter Physics, Institute of Physics, Chinese Academy of Sciences, Beijing 100190, People's Republic of China}

\author{Yujun Yang}
\affiliation{Institute of Atomic and Molecular Physics, Jilin University, Changchun 130012, People's Republic of China}

\author{Yingchun Guo}
\affiliation{School of Physics, East China Normal University, Shanghai 200241, People's Republic of China}

\author{Zhiyi Wei}
\affiliation{Beijing National Laboratory for Condensed Matter Physics, Institute of Physics, Chinese Academy of Sciences, Beijing 100190, People's Republic of China}
\affiliation{University of Chinese Academy of Sciences, Beijing 100049, People's Republic of China}
\affiliation{Songshan Lake Materials Laboratory, Dongguan 523808, People's Republic of China}

\author{Jing Chen}
\affiliation{Department of Modern Physics, and Hefei National Research Center for Physical Sciences at the Microscale and School of Physical Sciences, University of Science and Technology of China, Hefei 230026, People's Republic of China}

\author{Xiaojun Liu}
\affiliation{State Key Laboratory of Magnetic Resonance and Atomic and Molecular Physics, Wuhan Institute of Physics and Mathematics, Chinese Academy of Sciences, Wuhan 430071, People's Republic of China}

\author{Bingbing Wang}
\email{wbb@iphy.ac.cn}
\affiliation{Beijing National Laboratory for Condensed Matter Physics, Institute of Physics, Chinese Academy of Sciences, Beijing 100190, People's Republic of China}
\affiliation{University of Chinese Academy of Sciences, Beijing 100049, People's Republic of China}

\date{\today}

\begin{abstract}
Synthetic three-dimensional (3D) light fields provide a new route to imprinting chirality on achiral matter through light--matter interaction. Although previous work has numerically demonstrated that chirality can be imprinted on the photoelectron momentum spectrum of atoms, a transparent analytical physical picture that links the structure of the driving field to the chirality of the spectrum is still an open question. Here, using frequency-domain theory based on nonperturbative quantum electrodynamics (QED), we demonstrate that this chirality originates from coherence between adjacent quantum transition channels. We further find that a symmetry transformation introduces into the coherence term a symmetry-breaking phase shift ($\Phi_{\mathrm{SBPS}}$), which can be tuned by varying the laser-field parameters to preserve or break the corresponding symmetry in the spectrum. Once all such symmetries are broken, the momentum spectrum becomes chiral. In particular, for the noncollinear two-color 3D light field considered here, the spectrum is chiral for even $f=\omega_2/\omega_1$ and achiral for odd $f$. Our results provide a microscopic interpretation of chirality imprinting and establish a framework for predicting and controlling chiral electron wave packets with synthetic 3D light fields.

\end{abstract}

\maketitle

\section{Introduction}
\label{sec:introduction}
Chirality is a fundamental geometric property of three-dimensional objects: an object is chiral if it cannot be superimposed on its mirror image by translations and proper rotations, which is equivalent to the absence of any improper-rotation symmetry (including mirror and inversion symmetries)~\cite{Brandt2017}. Chirality appears widely in nature and plays a key role in biology, chemistry, and pharmacology~\cite{Peluso2022a}. Although the left and right enantiomers of a chiral molecule share the same chemical composition, they can exhibit strikingly different properties when interacting with other chiral systems~\cite{Agranat2002}.

Light has long been a powerful and precise tool for probing chiral molecules~\cite{Domingos2018,Janssen2014}. A prominent example is circular dichroism (CD)~\cite{Tang2010,Rouxel2024}, in which left- and right-handed enantiomers of randomly oriented chiral molecules are distinguished by their different absorption rates of circularly polarized light of one color. However, the CD signal usually relies on nondipole interactions and is therefore much weaker than other electric-dipole optical responses. The subsequently developed photoelectron circular dichroism (PECD)~\cite{Ordonez2019,Heger2025,Liu2025,Zhou2026,Zhou2026a} employs circularly polarized light to ionize chiral molecules and probes the dichroism in the angular distribution of the photoelectron rather than the total absorption intensity. This dichroism is significantly stronger because it already appears within the electric-dipole approximation~\cite{Bowering2001,Powis2008}. To further enhance such dichroic signals, researchers began to construct 3D light fields instead of 2D fields such as circularly polarized light. In 2019, Ayuso et al.~\cite{Ayuso2019} introduced a synthetic chiral light field based on noncollinear two-color beams, enabling randomly oriented chiral molecules to achieve electric-dipole chiral dichroism approaching its 200\% theoretical maximum and selectively quench one enantiomer. The idea of this synthetic chiral light field has since stimulated extensive research into its properties and its interactions with matter. On the one hand, subsequent studies have quantified the degree of chirality~\cite{Neufeld2020} and characterized its handedness~\cite{Neufeld2022}. On the other hand, synthetic chiral fields have been exploited to realize enantiosensitive light bending~\cite{Ayuso2021}, as well as strong chiral dichroism and enantiopurification~\cite{Neufeld2021}. Since then, light has been evolving from a tool for probing chirality into one for controlling chiral signals.

Furthermore, a natural question arises: can the chirality of light be used not only to probe and modulate the existing chirality of matter, but to directly imprint chirality on matter that is originally achiral? In 2022, Mayer et al.~\cite{Mayer2022} theoretically demonstrated that chirality can be imprinted on both the bound states of atoms and their photoelectron momentum spectra. Several works have further developed quantitative measures for such light-induced chirality~\cite{Katsoulis2022,Pisanty2026}. In 2025, Geyer et al.~\cite{Geyer2025} further showed theoretically that the chirality imprinted on the photoelectron momentum distribution of atoms can survive focal-volume integration under realistic experimental conditions, bringing this emerging field closer to experimental realization. Although light-induced chirality imprinting has been theoretically demonstrated in bound electronic chirality~\cite{Chen2024,Moitra2025,Pannir-Sivajothi2026,Liu2026}, in continuum electronic chirality and photoelectron momentum distributions~\cite{Mayer2022,Geyer2025}, and in the vibrational and rotational wave packets of achiral molecules~\cite{Leibscher2024,Blech2026}, a transparent analytical description of the underlying physical mechanism by which chirality is transferred from the light field to achiral matter is still an open question. 

In this work, we establish this connection using the frequency-domain theory based on nonperturbative quantum electrodynamics (QED). This theory was first proposed by Guo et al.~\cite{Guo1988,Guo1992a,Guo1992b} and has since been applied to a variety of strong-field processes, including above-threshold ionization (ATI)~\cite{Aberg1993}, high-order above-threshold ionization (HATI)~\cite{Wang2007}, and nonsequential double ionization (NSDI)~\cite{Wang2012,Jin2016}. Applying this method, we reveal that the chirality of the ATI spectrum of atoms originates from the coherent superposition of adjacent quantum transition channels, and we propose an analytical approach to selectively preserve or break the geometric symmetries of the 3D ionization spectrum by changing the laser conditions. We find that the coherence terms of the adjacent-channel superposition acquire a phase shift under certain symmetry transformations. By analogy with the symmetry-breaking phase shift that Ponomarev et al.~\cite{Ponomarev2009} discussed for the harmonic mixing signal of an ac-driven quantum motor, we name the corresponding phase of the coherence term as the symmetry-breaking phase shift, denoted by $\Phi_{\mathrm{SBPS}}$. By tuning the relevant field parameters, this phase can be switched between zero and nonzero values, thereby preserving or breaking the corresponding symmetry in the photoelectron spectrum. When all such symmetries are broken, the spectrum becomes chiral.

The paper is organized as follows. In Sec.~\ref{sec:theory}, we introduce the method used to calculate the ionization spectrum. In Sec.~\ref{sec:results}, we present the numerical results and summarize the rules governing the emergence of chiral and achiral spectra. In Sec.~\ref{sec:proof}, we provide an analytical proof of these rules. In Sec.~\ref{sec:discussion}, we show that the rules remain valid within the classical description of light as an electromagnetic wave. In Sec.~\ref{sec:conclusion}, we conclude the paper.
\FloatBarrier
\section{Theoretical Method}
\label{sec:theory}
\begin{figure}[H]
    \includegraphics[width=\columnwidth]{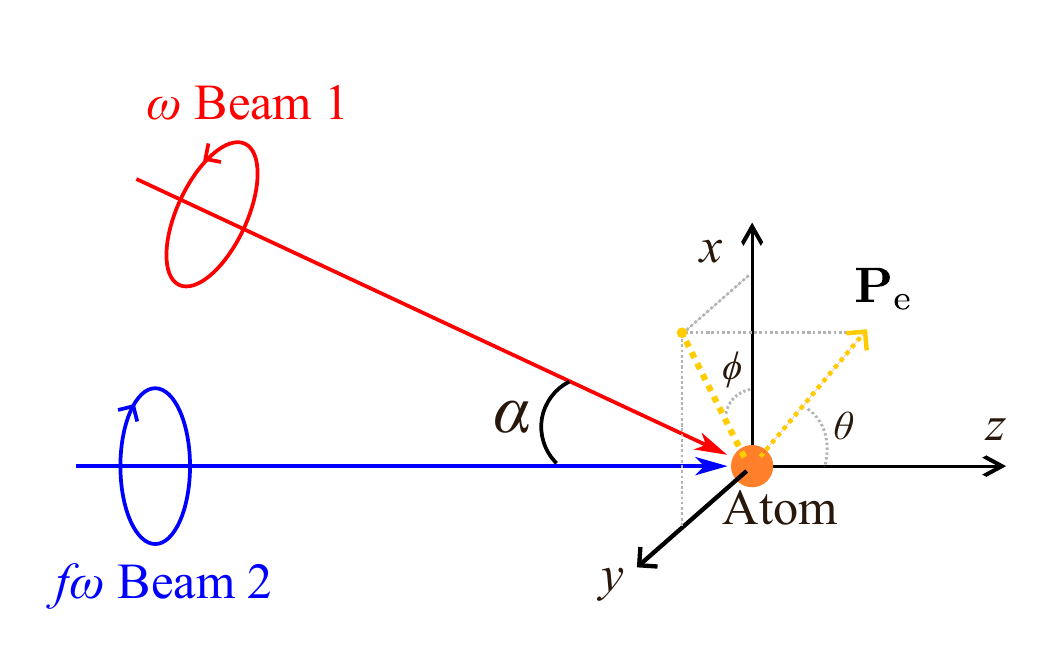}  
    \caption{\label{fig:1} Schematic of the synthetic 3D light field, composed of an elliptically polarized beam (Beam 1) at frequency $\omega$ and a circularly polarized beam (Beam 2) at frequency $f\omega$, intersecting at an angle $\alpha$.}
\end{figure}
Within the framework of the frequency-domain theory based on nonperturbative QED~\cite{Guo1992b}, the laser field is quantized and, together with the electron, forms a quantum state. During the ionization process, the combined electron--laser system is treated as an isolated system and satisfies energy conservation, so that formal scattering theory can be employed~\cite{Gell-Mann1953}. Here we consider only the direct ionization process. The transition matrix element from the initial state $|\Psi_{\mathrm{i}}\rangle$ to the final state $|\Psi_{\mathrm{f}}\rangle$ can be written as~\cite{Gell-Mann1953}
\begin{equation}
T_{\mathrm{ATI}}=\langle\Psi_{\mathrm{f}}|V|\Psi_{\mathrm{i}}\rangle,
\label{eq:transition}
\end{equation}
where the electron--field interaction potential $V$ takes the form~\cite{Guo1992b}
\begin{equation}
V=-\frac{e}{2m_{\mathrm{e}}}\big[(-i\nabla)\cdot\mathbf{A}+\mathbf{A}\cdot(-i\nabla)\big]+\frac{e^{2}}{2m_{\mathrm{e}}}\mathbf{A}^{2},
\label{eq:potential}
\end{equation}
with $\mathbf{A}(\mathbf{r})=\mathbf{A}_{1}(\mathbf{r})+\mathbf{A}_{2}(\mathbf{r})$ the total vector potential where $\mathbf{A}_{s}(\mathbf{r})=g_{s}(\hat{\epsilon}_{s}e^{i\mathbf{k}_{s}\cdot\mathbf{r}}a_{s}+\mathrm{H.c.})$, and $g_{s}=(2\omega_{s}V_{\mathrm{norm}})^{-1/2}$ for $s=1,2$, $V_{\mathrm{norm}}$ is the normalization volume of the laser field, and $\hat{\epsilon}_{s}$ is the normalized polarization vector. In this work, we consider a noncollinear two-color circular--elliptical polarized field~\cite{Mayer2022}, as shown in Fig.~\ref{fig:1}. Beam~1 is at the fundamental frequency $\omega$ ($\lambda = 800$~nm), and Beam~2 is at frequency $f\omega$, with the intensities of both beams are $31$~TW/cm$^{2}$. To simplify the calculations, we adjust the ellipticity of Beam~1 so that its projection onto the $xy$ plane is circularly polarized. The polarization vectors of the two beams can then be written as
\begin{equation}
\begin{aligned}
\hat{\epsilon}_{1} & = \frac{1}{\sqrt{1+\cos^{2}\alpha}}(\cos\alpha,i\cos\alpha,\sin\alpha)e^{i\phi_{\mathrm{A}1}},\\
\hat{\epsilon}_{2} & = \frac{1}{\sqrt{2}}(1,-i,0)e^{i\phi_{\mathrm{A}2}}.
\end{aligned}
\label{eq:polarization}
\end{equation}
Since $\mathbf{E}=-\partial\mathbf{A}/\partial t$, the initial phase of the vector potential $\phi_{\mathrm{A}}$ and that of the electric field $\phi_{\mathrm{E}}$ satisfy $\phi_{\mathrm{E}}=-\phi_{\mathrm{A}}-\pi/2$. In this work, we set $\phi_{\mathrm{E}1}=\phi_{\mathrm{E}2}=0$, i.e., $\phi_{\mathrm{A}1}=\phi_{\mathrm{A}2}=-\pi/2$. The initial state can be written as~\cite{Li2024,Li2025} $|\Psi_{\mathrm{i}}\rangle=\Phi_{\mathrm{i}}(\mathbf{r})\otimes|l_{1}\rangle\otimes|l_{2}\rangle$, with energy $E_{\mathrm{i}}=-E_{\mathrm{B}}+(l_{1}+\frac{1}{2})\omega_{1}+(l_{2}+\frac{1}{2})\omega_{2}$, where $\Phi_{\mathrm{i}}(\mathbf{r})$ is the bound state of the electron with binding energy $E_{\mathrm{B}}$, and $|l_{s}\rangle$ is the Fock state with photon number $l_{s}$. The final state is given by the Volkov state~\cite{Guo1992b} $|\Psi_{\mathrm{f}}\rangle=|\Psi_{\mathbf{P}_{\mathrm{e}}m_{1}m_{2}}\rangle$, which is the eigenstate of a free electron moving in the two-color laser field,
\begin{equation}
\begin{aligned}
|\Psi_{\mathbf{P}_{\mathrm{e}}m_{1}m_{2}}\rangle={}&V_{\mathrm{norm}}^{-1/2}\sum_{\substack{q_{1}=-m_{1}\\q_{2}=-m_{2}}}^{\infty}e^{i\mathbf{P}_{\mathrm{e}}\cdot\mathbf{r}}e^{i(u_{p_{1}}-q_{1})\mathbf{k}_{1}\cdot\mathbf{r}}\\
&\times e^{i(u_{p_{2}}-q_{2})\mathbf{k}_{2}\cdot\mathbf{r}}\mathscr{J}_{q_{1},q_{2}}(\zeta)|m_{1}+q_{1},m_{2}+q_{2}\rangle,
\end{aligned}
\label{eq:vlf}
\end{equation}
with energy $E_{\mathbf{P}_{\mathrm{e}}m_{1}m_{2}}=\frac{1}{2}P_{\mathrm{e}}^{2}+(m_{1}+\frac{1}{2})\omega_{1}+(m_{2}+\frac{1}{2})\omega_{2}+u_{p_{1}}\omega_{1}+u_{p_{2}}\omega_{2}$, $q_{s}=l_{s}-m_{s}$ is the number of photons absorbed from the laser field, $u_{p_{s}}$ is the ponderomotive energy in units of the photon energy, $\mathbf{k}_{s}$ is the photon momentum of the laser field.

The generalized Bessel function $\mathscr{J}_{q_{1},q_{2}}(\zeta)$ for our field configuration takes the form
\begin{equation}
\begin{aligned}
\mathscr{J}_{q_{1},q_{2}}(\zeta)={}&\sum_{q_{3},q_{5}}J_{-q_{1}+2q_{3}+q_{5}}(\zeta_{1})\,e^{i(-q_{1}+2q_{3}+q_{5})\phi_{\mathbf{P}_{\mathrm{e}},1}}\\
&\times J_{-q_{2}+q_{5}}(\zeta_{2})\,e^{i(-q_{2}+q_{5})\phi_{\mathbf{P}_{\mathrm{e}},2}}\\
&\times J_{-q_{3}}(\zeta_{3})J_{-q_{5}}(\zeta_{5})\,e^{-i(q_{1}\phi_{\mathrm{A}1}+q_{2}\phi_{\mathrm{A}2})},
\end{aligned}
\label{eq:bessel}
\end{equation}
where $J_{n}(x)$ is the Bessel function of order $n$, $q_{3}$ and $q_{5}$ are integers, and $\phi_{\mathbf{P}_{\mathrm{e}},s}$ is the phase arising from the momentum-dependent term.
The parameters in Eq.~(\ref{eq:bessel}) can be expressed as
\begin{equation}
\begin{aligned}
\zeta_1\ &=\ \frac{2P_{\mathrm{e}}}{\sqrt{1+\cos^2\alpha}}\sqrt{\frac{u_{p1}}{\omega_1}}
\left|\sin\alpha\cos\theta+\cos\alpha\sin\theta\,e^{i\phi}\right|,\\
\phi_{\mathbf{P}_{\mathrm{e}},1}\ &=\ \arg\!\big(\sin\alpha\cos\theta+\cos\alpha\sin\theta\,e^{i\phi}\big),\\[4pt]
\zeta_2\ &=\ P_\mathrm{e}\sqrt{\frac{2u_{p2}}{\omega_2}}\sin\theta,
\qquad \phi_{\mathbf{P}_{\mathrm{e}},2}=\arg(\sin\theta e^{-i\phi}),\\[4pt]
\zeta_3\ &=\ \frac{u_{p1}}{2}\frac{\sin^2\alpha}{1+\cos^2\alpha},
\qquad \phi_3=2\phi_{\mathrm{A}1},\\[4pt]
\zeta_5\ &=\ \frac{2\sqrt{2f\,u_{p1}u_{p2}}\cos\alpha}{(1+f)\sqrt{1+\cos^2\alpha}},
\qquad \phi_5=\phi_{\mathrm{A}1}+\phi_{\mathrm{A}2}.
\end{aligned}
\label{eq:zeta}
\end{equation}
Under the dipole approximation, the ATI transition matrix element can be written as~\cite{Li2024}
\begin{multline}
T_{q_1,q_2}=V_{\mathrm{norm}}^{-1/2}\left[\left(u_{p_{1}}-q_{1}\right)\omega_{1}
+\left(u_{p_{2}}-q_{2}\right)\omega_{2}\right]\\
\times\mathscr{J}_{q_1,q_2}(\zeta)\Phi_{\mathrm{i}}\left(\mathbf{P}_{\mathrm{e}}\right),
\label{eq:Tq1q2}
\end{multline}
in which the single-channel relation should be satisfied: $q_1\omega_1+q_2\omega_2-u_{p_1}\omega_1-u_{p_2}\omega_2-E_\mathrm{B}=\frac{1}{2}P_{\mathrm{e}}^{2}.$ We map each channel onto the three-dimensional momentum space and sum over all channels to obtain the relative intensity of the ionization momentum spectrum:
\begin{equation}
\begin{aligned}
W(\mathbf{P}_{\mathrm{e}})\propto P_\mathrm{e} \Bigg|\sum_{(q_1,q_2)}T_{q_1,q_2}\Bigg|^{2}.
\label{eq:Wp}
\end{aligned}
\end{equation}

\FloatBarrier
\section{Numerical Results}
\label{sec:results}

\begin{figure*}[t]
    \centering
    \includegraphics[width=0.8\textwidth]{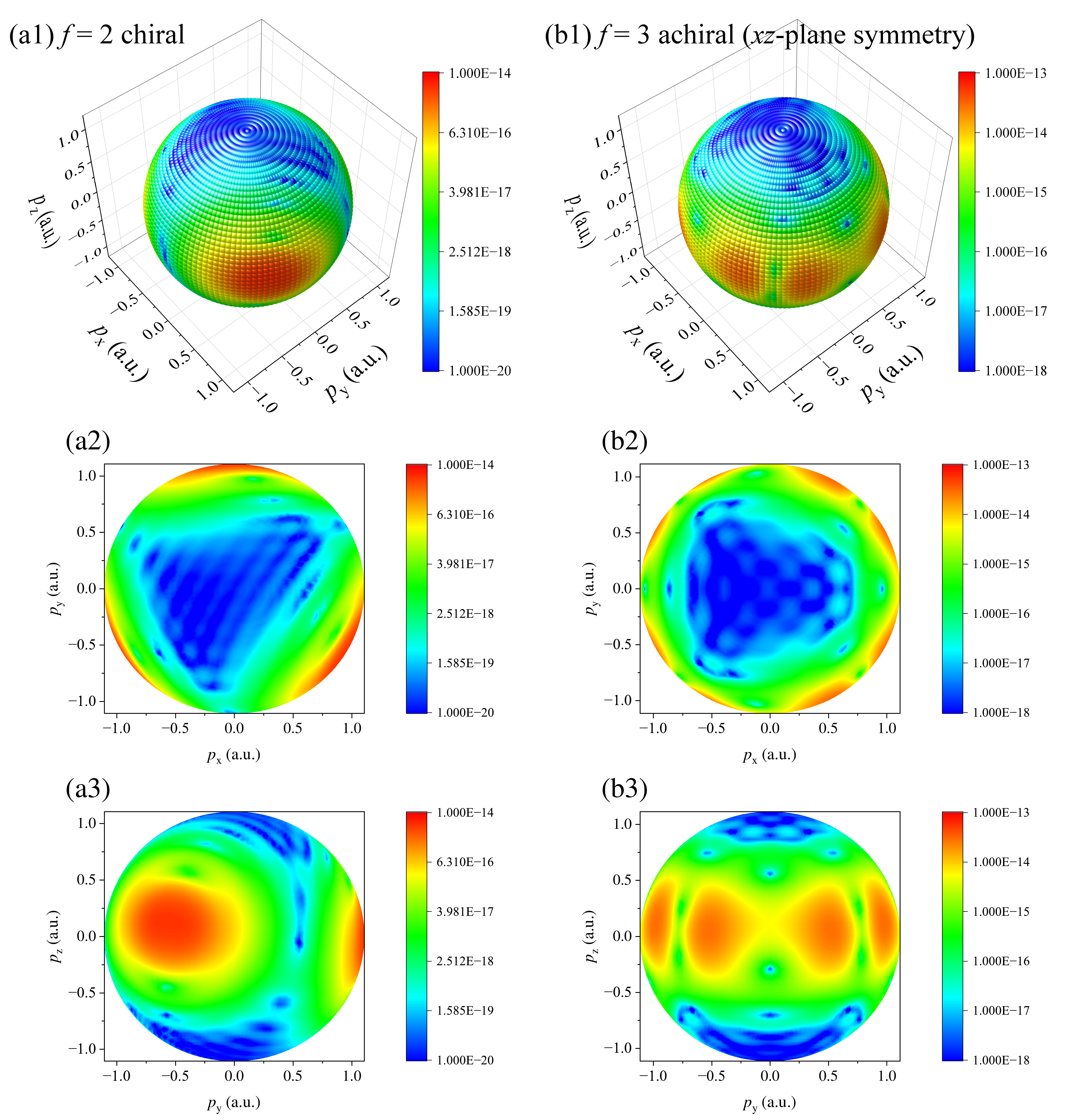}  
    \caption{Photoelectron momentum distributions on the spherical shell at $E = 1.105$ a.u. with a fixed crossing angle $\alpha=10^\circ$, for the case of $f=2$ (a1) and its top (a2) and side (a3) projections; the corresponding results for $f=3$ are shown in (b1)–(b3). Unless otherwise stated, atomic units are used throughout.}
    \label{fig:2}
\end{figure*}

To visualize whether the spectrum is chiral or achiral, we display in Fig.~\ref{fig:2} the spherical-shell photoelectron momentum distributions of the emitted photoelectrons at the fixed energy $E=1.105$ a.u. As shown in Fig.~\ref{fig:2}, the spectrum for $f=2$ (Figs.~\hyperref[fig:2]{\ref*{fig:2}(a1)}--\hyperref[fig:2]{\ref*{fig:2}(a3)}) is chiral and exhibits no geometric symmetry, whereas the spectrum for $f=3$ (Figs.~\hyperref[fig:2]{\ref*{fig:2}(b1)}--\hyperref[fig:2]{\ref*{fig:2}(b3)}) is achiral and possesses symmetry with respect to the $xz$ plane. To more intuitively reveal the chiral features of the photoelectron ionization spectrum, we follow the definition of the overlap measure in Ref.~\cite{Mayer2022} and quantify the chirality of the spectrum by 
\begin{equation}
\begin{aligned}
\chi &= \frac{\min_{\rho}\int \mathrm{d}^3\mathbf{P}_{\mathrm{e}}\; \left|W_M^{\rho}(\mathbf{P}_{\mathrm{e}}) - W(\mathbf{P}_{\mathrm{e}})\right|}
{\int \mathrm{d}^3\mathbf{P}_{\mathrm{e}}\; W(\mathbf{P}_{\mathrm{e}})},
\end{aligned}
\label{eq:chi}
\end{equation}
where $W(\mathbf{P}_{\mathrm{e}})$ is the original spectrum under test, and $W_M^{\rho}(\mathbf{P}_{\mathrm{e}})$ is obtained from $W(\mathbf{P}_{\mathrm{e}})$ by mirror reflection through the $yz$ plane followed by rotations by the Euler angles $\rho = (\alpha_{\rho}, \beta_{\rho}, \gamma_{\rho})_{zyz}$. The minimization $\min_{\rho}$ is performed over all possible three-dimensional rotations, i.e., over all sets of Euler angles, to find the best overlap between $W(\mathbf{P}_{\mathrm{e}})$ and its mirror-reflected counterpart. A value of $\chi=0$ indicates an achiral spectrum, whereas $\chi \neq 0$ indicates a chiral spectrum. We have computed $\chi$ and the corresponding optimal overlap (Euler) angle for frequency ratios $f=2$--$7$, and the results are listed in Table~\ref{tab:chi}, where the value of $\chi$ is evaluated from the full three-dimensional momentum spectrum by integrating over the entire momentum ball from $E=0$ up to $E=50\omega$. 
\begin{table}[t]
\centering
\caption{Chirality measure $\chi$ and optimal overlap Euler angles $\rho=(\alpha_\rho,\beta_\rho,\gamma_\rho)_{zyz}$ for different frequency ratios $f$ at a fixed crossing angle $\alpha=10^\circ$.}
\label{tab:chi}
\begin{ruledtabular}
\begin{tabular}{ccccc}
$f$ & $\chi$ & \multicolumn{2}{c}{$\rho=(\alpha_\rho,\beta_\rho,\gamma_\rho)_{zyz}$ ($^\circ$)} & Chirality \\
\colrule
2 & $1.16\times10^{-1}$ & \multicolumn{2}{c}{$(240.05, 173.48, 119.81)$} & Chiral \\
3 & $1.97\times10^{-15}$ & \multicolumn{2}{c}{$(180.00, 0.00, 0.00)$} & Achiral \\
4 & $7.54\times10^{-2}$ & \multicolumn{2}{c}{$(72.49, 3.70, 72.12)$} & Chiral \\
5 & $1.26\times10^{-15}$ & \multicolumn{2}{c}{$(180.00, 0.00, 0.00)$} & Achiral \\
6 & $3.35\times10^{-2}$ & \multicolumn{2}{c}{$(102.80, 177.25, 257.30)$} & Chiral \\
7 & $8.73\times10^{-16}$ & \multicolumn{2}{c}{$(180.00, 0.00, 0.00)$} & Achiral \\
\end{tabular}
\end{ruledtabular}
\end{table}
As shown in Table~\ref{tab:chi}, one may find that the spectrum is chiral ($\chi \neq 0$) for even frequency ratios $f$, and achiral ($\chi \approx 0$) for odd $f$. Additionally, for even $f$, we find that $\chi$ decreases as $f$ increases, and the Euler angles $\rho=(\alpha_\rho,\beta_\rho,\gamma_\rho)_{zyz}$ differ among different values of $f$. For odd $f$, the original spectrum coincides with its mirror reflection about the $yz$ plane after a rotation by $\rho=(180^\circ,0^\circ,0^\circ)_{zyz}$, indicating that the original spectrum possesses $xz$-plane symmetry which is also visible in Figs.~\hyperref[fig:2]{\ref*{fig:2}(b1)}--\hyperref[fig:2]{\ref*{fig:2}(b3)}. We will provide an analytical proof of these rules in Sec.~\ref{sec:proof}. 

\begin{figure}[t]
    \centering
    \includegraphics[width=\columnwidth]{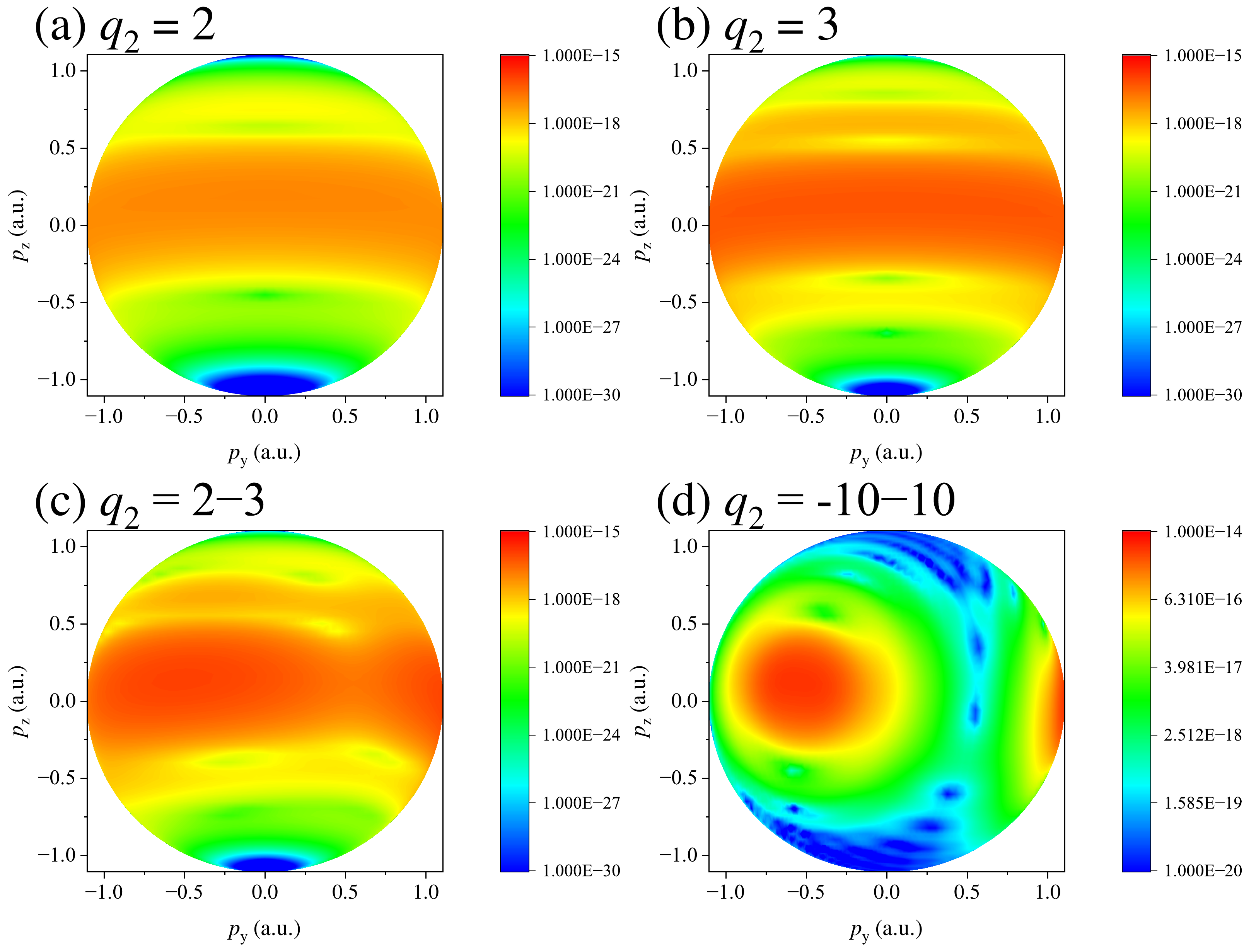}  
    \caption{Photoelectron momentum distributions on the spherical shell at $E = 1.105$ a.u.\ for the case of $f=2$: single-channel side projections for $q_2=2$ (a) and $q_2=3$ (b); the coherent superposition of these two channels (c); and the multi-channel superposition with $q_2$ ranging from $-10$ to $10$ (d). The distribution in (d) is already close to the all-channel result shown in Fig.~\hyperref[fig:2]{\ref*{fig:2}(a3)}.}
    \label{fig:4}
\end{figure}

To investigate the mechanism underlying chirality imprinting on the spectrum by chiral light fields, we adopt the framework of quantum transition channels.
\begin{figure}[t]
    \includegraphics[width=\columnwidth]{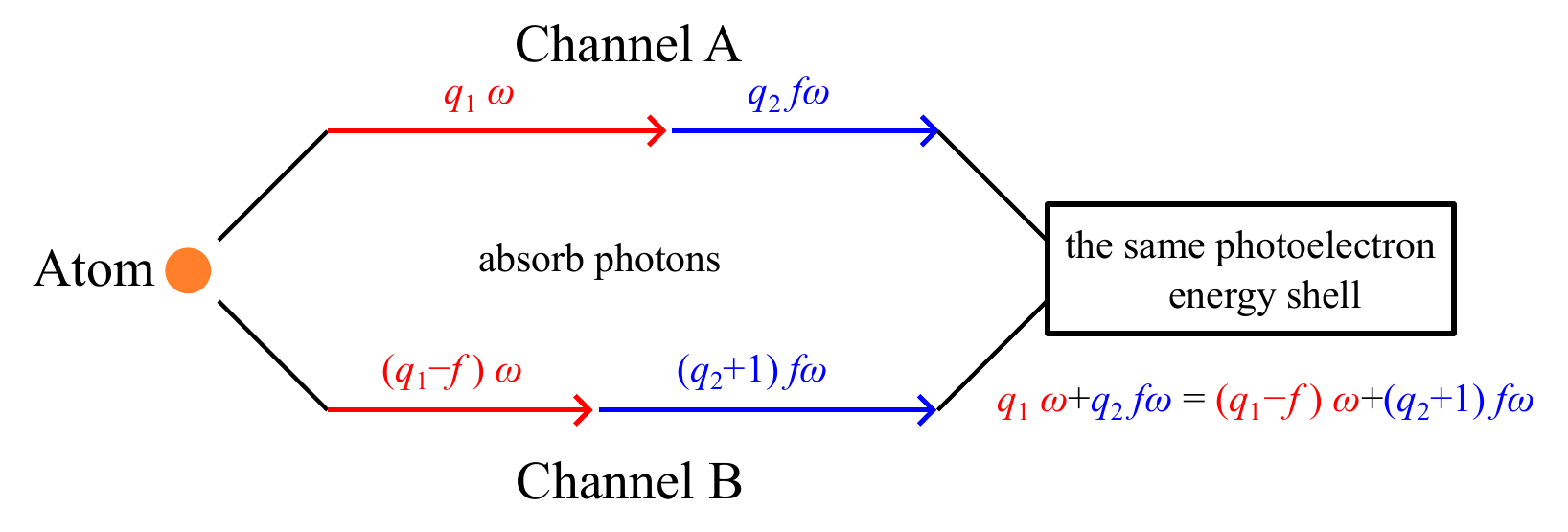}  
    \caption{Schematic of quantum transition channels (shown for two adjacent channels). Channel B absorbs $f$ fewer $\omega$ photons and one additional $f\omega$ photon compared to channel A. The two channels end on the same photoelectron energy shell and exhibit coherence.}
    \label{fig:3}
\end{figure}
The entire 3D momentum spectrum is composed of spherical shells at different energies, and each energy shell is formed by the coherent superposition of different quantum transition channels, which is labeled by $q_2$~\cite{Li2025}, as shown in Fig.~\ref{fig:3}. To elucidate the origin of the chirality, without loss of generality, we take channels $q_2=2$ and $q_2=3$ for the case of $f=2$ as examples, and calculate the spherical-shell spectra for these two single channels (Figs.~\hyperref[fig:4]{\ref*{fig:4}(a)}
and
\hyperref[fig:4]{\ref*{fig:4}(b)}), their coherent superposition (Fig.~\hyperref[fig:4]{\ref*{fig:4}(c)}), and the multi-channel superposition (Fig.~\hyperref[fig:4]{\ref*{fig:4}(d)}).
Figure~\ref{fig:4} shows that the single-channel spectra are symmetric with respect to the $xz$ plane, whereas the two-channel and multi-channel superpositions break this symmetry and become chiral. This demonstrates that the chirality stems from the coherent superposition of adjacent quantum transition channels, as we will prove analytically in Sec.~\ref{sec:proof}.
\FloatBarrier

\section{Analytical Proof}
\label{sec:proof}

To analyze the origin of the chirality, we first consider a single channel. According to previous work~\cite{Li2025}, the main features of the atomic momentum spectrum originate from the generalized Bessel function, which also contains the angular variables $\theta$ and $\phi$ involved in the symmetry transformations. Without loss of generality, from Eqs.~(\ref{eq:bessel}), (\ref{eq:zeta}) and (\ref{eq:Wp}), the possible symmetries that a single-channel ionization spectrum may retain are the $xz$-plane reflection symmetry ($\hat{\sigma}_{xz}: \phi \to -\phi$) and the spatial inversion symmetry ($\hat{\mathcal{P}}: \theta \to \pi - \theta, \phi \to \pi + \phi$). Therefore, by using Eq.~(\ref{eq:bessel}), we have
\begin{equation}
\begin{aligned}
\hat{\sigma}_{xz}\left|\mathscr{J}_{q_1,q_2}(\zeta)\right|^2 
&= \left|\mathscr{J}_{q_1,q_2}^{*}(\zeta)\right|^2 
= \left|\mathscr{J}_{q_1,q_2}(\zeta)\right|^2, \\
\text{and}\quad
\hat{\mathcal{P}}\left|\mathscr{J}_{q_1,q_2}(\zeta)\right|^2 
&= \left|\mathscr{J}_{q_1,q_2}(\zeta)e^{-i\pi(q_1+q_2)}\right|^2 \\
&= \left|\mathscr{J}_{q_1,q_2}(\zeta)\right|^2,
\end{aligned}
\label{eq:symm}
\end{equation}
hence, the single-channel spectrum remains invariant under both 
$\hat{\sigma}_{xz}$ and $\hat{\mathcal{P}}$, i.e., the single-channel spectrum 
possesses both the $xz$-plane reflection symmetry and the spatial inversion symmetry.

Next, we consider the coherent superposition of channel $(q_1,q_2)$ and its adjacent channel $(q_1-f,q_2+1)$. As shown in Fig.~\ref{fig:3}, these two channels share the same final-state energy and can therefore exhibit coherence. For mathematical convenience, we define
\begin{subequations}\label{eq:gdef}
\begin{equation}
\begin{split}
g_{q_1,q_2}(\zeta) ={}& \sum_{q_3,q_5}
J_{-q_1+2q_3+q_5}(\zeta_1)
J_{-q_2+q_5}(\zeta_2)\\
&\times J_{-q_3}(\zeta_3)J_{-q_5}(\zeta_5)
e^{i(2q_3+q_5)\phi_{\mathbf{P}_{\mathrm{e}},1}}
e^{iq_5\phi_{\mathbf{P}_{\mathrm{e}},2}},
\end{split}
\end{equation}
\begin{equation}
\begin{split}
g_{q_1-f,q_2+1}(\zeta) ={}& \sum_{q_3,q_5}
J_{-(q_1-f)+2q_3+q_5}(\zeta_1)
J_{-(q_2+1)+q_5}(\zeta_2)\\
&\times J_{-q_3}(\zeta_3)J_{-q_5}(\zeta_5)
e^{i(2q_3+q_5)\phi_{\mathbf{P}_{\mathrm{e}},1}}
e^{iq_5\phi_{\mathbf{P}_{\mathrm{e}},2}},
\end{split}
\end{equation}
\begin{equation}
\begin{split}
\text{and}\quad \mathscr{G}(\zeta) ={}&
g_{q_1,q_2}(\zeta)
g_{q_1-f,q_2+1}^{*}(\zeta)
e^{-if\phi_{\mathbf{P}_{\mathrm{e}},1}}
e^{i\phi_{\mathbf{P}_{\mathrm{e}},2}}\\
&\times e^{-i(f\phi_{\mathrm{A}1}-\phi_{\mathrm{A}2})}.
\end{split}
\end{equation}
\end{subequations}
We thus obtain
\begin{subequations}\label{eq:channelsum}
\begin{equation}
\begin{split}
&\left|\mathscr{J}_{q_1,q_2}(\zeta)+\mathscr{J}_{q_1-f,q_2+1}(\zeta)\right|^2\\
&\quad=\left|\mathscr{J}_{q_1,q_2}(\zeta)\right|^2
+\left|\mathscr{J}_{q_1-f,q_2+1}(\zeta)\right|^2\\
&\quad\quad+\mathscr{G}(\zeta)+\mathscr{G}^{*}(\zeta),
\end{split}
\end{equation}
\begin{equation}
\begin{split}
\text{with}\quad
&\hat{\sigma}_{xz}\left|\mathscr{J}_{q_1,q_2}(\zeta)+\mathscr{J}_{q_1-f,q_2+1}(\zeta)\right|^2\\
&\quad=\left|\mathscr{J}_{q_1,q_2}(\zeta)\right|^2
+\left|\mathscr{J}_{q_1-f,q_2+1}(\zeta)\right|^2\\
&\quad\quad+\mathscr{G}(\zeta)e^{i2(f\phi_{\mathrm{A}1}-\phi_{\mathrm{A}2})}\\
&\quad\quad+\mathscr{G}^{*}(\zeta)e^{-i2(f\phi_{\mathrm{A}1}-\phi_{\mathrm{A}2})},
\end{split}
\end{equation}
\begin{equation}
\begin{split}
\text{and}\quad
&\hat{\mathcal{P}}\left|\mathscr{J}_{q_1,q_2}(\zeta)+\mathscr{J}_{q_1-f,q_2+1}(\zeta)\right|^2\\
&\quad=\left|\mathscr{J}_{q_1,q_2}(\zeta)\right|^2
+\left|\mathscr{J}_{q_1-f,q_2+1}(\zeta)\right|^2\\
&\quad\quad+\mathscr{G}(\zeta)e^{i\pi(-f+1)}\\
&\quad\quad+\mathscr{G}^{*}(\zeta)e^{i\pi(f-1)}.
\end{split}
\end{equation}
\end{subequations}
It is evident that, under these two symmetry operations, the coherence terms each acquire an additional phase. Following the terminology introduced by Ponomarev et al.~\cite{Ponomarev2009} for the harmonic mixing signal of an ac-driven quantum motor, we name this phase as the symmetry-breaking phase shift of the coherence term, $\Phi_{\mathrm{SBPS}}$:
\begin{equation}
\begin{aligned}
\Phi_{\mathrm{SBPS}}(\hat{\sigma}_{xz})&=2(f\phi_{\mathrm{A}1}-\phi_{\mathrm{A}2}),\\
\text{and}\quad \Phi_{\mathrm{SBPS}}(\hat{\mathcal{P}})&=\pi(-f+1).
\end{aligned}
\label{eq:SBPS}
\end{equation}
For the final spectrum to be chiral, both symmetries must be broken, i.e., the phase shifts must render the transformed coherence terms distinct from their original forms. The light field therefore needs to satisfy
\begin{equation}
\begin{aligned}
\Phi_{\mathrm{SBPS}}(\hat{\sigma}_{xz}) &\neq 2n\pi, \\
\text{and}\quad \Phi_{\mathrm{SBPS}}(\hat{\mathcal{P}}) &\neq 2m\pi, \quad n,m\in\mathbb{Z}.
\end{aligned}
\label{eq:condition}
\end{equation}
For $f=2$, $\Phi_{\mathrm{SBPS}}(\hat{\sigma}_{xz})=\Phi_{\mathrm{SBPS}}(\hat{\mathcal{P}})=-\pi$, so the spectrum is chiral; for $f=3$, $\Phi_{\mathrm{SBPS}}(\hat{\sigma}_{xz})=\Phi_{\mathrm{SBPS}}(\hat{\mathcal{P}})=-2\pi$, so the spectrum is achiral. These conditions correspond exactly to the rules presented in Fig.~\ref{fig:2} and Table~\ref{tab:chi}.

Our study addresses the question of what light field can produce a chiral photoelectron spectrum by providing a new perspective: the symmetry-breaking constraints may be derived directly by analyzing the coherence term of adjacent-channel superposition. In particular, the symmetry transformation introduces a field-dependent symmetry-breaking phase shift $\Phi_{\mathrm{SBPS}}$ in the coherence term. Tuning field parameters controls this phase shift, selectively preserving or breaking spectral symmetries and thereby imprinting the light field's geometry onto the ionization spectrum.

\section{Discussion}
\label{sec:discussion}

To elucidate the classical counterpart of the above quantum description, we now consider the corresponding classical driving field. The electric field vector of our synthetic 3D light field can be written as~\cite{Mayer2022}
\begin{equation}
\begin{aligned}
E_{x}={}&E_{\omega}\cos(\omega t+\phi_{\mathrm{E}1})\cos\alpha\\
&+E_{f\omega}\cos(f\omega t+\phi_{\mathrm{E}2}),\\
E_{y}={}&\cos\alpha\,E_{\omega}\sin(\omega t+\phi_{\mathrm{E}1})\\
&-E_{f\omega}\sin(f\omega t+\phi_{\mathrm{E}2}),\\
E_{z}={}&E_{\omega}\cos(\omega t+\phi_{\mathrm{E}1})\sin\alpha,
\end{aligned}
\label{eq:efield}
\end{equation}
where $\phi_{\mathrm{E}1}$ and $\phi_{\mathrm{E}2}$ are the initial phases of the electric fields of the two beams. For classical fields, the chirality can be determined from the chiral field correlation functions. Following the work in Ref.~\cite{Ayuso2019}, the chiral field correlation function in the frequency domain reads
\begin{equation}
\begin{gathered}
h^{(2N+1)} = \mathbf{F}_{1}\cdot \left(\mathbf{F}_{2}\times\mathbf{F}_{3}\right)
\prod_{j=1}^{N-1} \left( \mathbf{F}_{2j+2}\cdot\mathbf{F}_{2j+3} \right), \\[2pt]
\omega_{1}+\omega_{2}+\cdots+\omega_{2N+1}=0, \qquad N\in\mathbb{Z}^{+},
\end{gathered}
\label{eq:hpseudo}
\end{equation}
where $\mathbf{F}_{i}$ is the Fourier component of the electric field vector at frequency $\omega_{i}$. Its time-domain counterpart is obtained by averaging the pseudoscalar formed by the electric field over time, so as to sample the entire Lissajous curve:
\begin{equation}
\begin{aligned}
&H^{(2N+1)}(\tau_1,\tau_2,\ldots,\tau_{2N})\\
&\quad=\int \mathrm{d}t\, \mathbf{E}(t)\cdot
\left[ \mathbf{E}(t+\tau_1) \times \mathbf{E}(t+\tau_2) \right]\\
&\quad\quad\times \prod_{j=1}^{N-1} \left[ \mathbf{E}(t+\tau_{2j+1}) \cdot \mathbf{E}(t+\tau_{2j+2}) \right],\\
&\quad\quad\quad N\in\mathbb{Z}^{+}.
\end{aligned}
\label{eq:Hpseudo}
\end{equation}
To determine whether a light field is chiral, one can construct the chiral field correlation functions order by order, starting from the lowest-order term $h^{(3)}$. If any order yields a nonzero correlation function, the field is chiral; if all orders are strictly zero, the field is achiral.

We first prove that the field is achiral when $f$ is odd. Applying a half-cycle translation of the fundamental frequency, $t \to t+T/2$ with $T/2=\pi/\omega$, the field components transform as $\mathbf{E}_{\omega}(t+T/2)=-\mathbf{E}_{\omega}(t)$ and $\mathbf{E}_{f\omega}(t+T/2)=(-1)^{f}\mathbf{E}_{f\omega}(t)$. For odd $f$, both components flip sign, and the shifted correlation function reads $H_{\mathrm{shifted}}^{(2N+1)}=(-1)^{2N+1}H^{(2N+1)}=-H^{(2N+1)}$.
On the other hand, the full-cycle average is independent of the time origin, so the same correlation function must remain unchanged:
\begin{equation}
H_{\mathrm{shifted}}^{(2N+1)}=H^{(2N+1)}=0.
\label{eq:Hshift2}
\end{equation}
Therefore, when $f$ is odd, the chiral field correlation functions vanish at all orders, and the light field is achiral.

We now prove that the light field is chiral when $f$ is even. First, we perform the Fourier decomposition of the two-color field in Eq.~(\ref{eq:efield}). In complex exponential representation, the electric field can be written as
\begin{equation}
\mathbf{E}(t)=\mathbf{F}(\omega)e^{-i\omega t}+\mathbf{F}(f\omega)e^{-if\omega t}+\mathrm{c.c.},
\label{eq:fourier}
\end{equation}
where
\begin{equation}
\begin{aligned}
\mathbf{F}(\omega) &= \frac{E_\omega}{2} e^{-i\phi_{\mathrm{E}1}} \left( \cos\alpha,\, i\cos\alpha,\, \sin\alpha \right),\\
\mathbf{F}(-\omega) &= \frac{E_\omega}{2} e^{i\phi_{\mathrm{E}1}} \left( \cos\alpha,\, -i\cos\alpha,\, \sin\alpha \right),\\
\mathbf{F}(f\omega) &= \frac{E_{f\omega}}{2} e^{-i\phi_{\mathrm{E}2}} \left( 1,\,-i,\,0 \right),\\
\mathbf{F}(-f\omega) &= \frac{E_{f\omega}}{2} e^{i\phi_{\mathrm{E}2}} \left( 1,\,i,\,0 \right).
\end{aligned}
\label{eq:fcomps}
\end{equation}
Using Eqs.~(\ref{eq:hpseudo}) and (\ref{eq:fourier}), we construct the chiral field correlation function
\begin{equation}
h^{(f+3)}=\left\{\mathbf{F}(-f\omega)\cdot[\mathbf{F}(-\omega)\times\mathbf{F}(\omega)]\right\}\left[\mathbf{F}(\omega)\cdot\mathbf{F}(\omega)\right]^{f/2},
\label{eq:hfplus3}
\end{equation}
which satisfies the condition $\omega_1+\cdots+\omega_{2N+1}=0$ in Eq.~(\ref{eq:hpseudo}). Substituting Eq.~(\ref{eq:fcomps}) into Eq.~(\ref{eq:hfplus3}) yields
\begin{equation}
h^{(f+3)}=-\frac{i}{2^{f+2}}E_{\omega}^{f+2}E_{f\omega}\cos\alpha(\sin\alpha)^{f+1}\, e^{i(\phi_{\mathrm{E}2}-f\phi_{\mathrm{E}1})}.
\label{eq:hresult}
\end{equation}
Therefore, when $f$ is even, $h^{(f+3)}\neq 0$ in the general case of $\sin\alpha\cos\alpha\neq 0$, and the light field is chiral.

The above analysis demonstrates that the classical method yields the same rule as our frequency-domain theory in Sec.~\ref{sec:proof}, but its judgment procedure is relatively complicated. The advantage of our frequency-domain theory is that it directly and analytically preserves symmetry or chirality selectively by tuning $\Phi_{\mathrm{SBPS}}$.

\section{Conclusion}
\label{sec:conclusion}
In summary, using the frequency-domain theory based on nonperturbative QED, we have shown that chirality imprinting by 3D light fields on ATI spectra originates from coherence between adjacent quantum transition channels. We proposed a method to control the spectral symmetry via a symmetry-breaking phase shift $\Phi_{\mathrm{SBPS}}$: by tuning the field parameters, one can selectively preserve or break the relevant symmetries, and chirality is imprinted when all such symmetries are broken. For the light field designed here, odd $f$ retains the $xz$-plane reflection and spatial inversion symmetries, giving an achiral ATI spectrum; while even $f$ breaks them and yields a chiral spectrum. This even-odd rule agrees with the classical chirality condition. Our analytical and controllable approach points toward designing and manipulating chirality in achiral matter.

\begin{acknowledgments}
We thank all the members of the SFAMP club for helpful discussions. This work was supported by the National Natural Science Foundation of China (Grants No. 92250303, No. 12474288, No. 12204526, No. 12104285, No. 11334009, and No. 11425414), the National Key R\&D Program of China (Grant No. 2022YFA1604200), the Guangdong Basic and Applied Basic Research Foundation (Grant No. 2022A1515011742), the Innovation Capability Support Program of Shaanxi Province (Grant No. 2024ZC-KJXX048), and the Shantou University Scientific Research Initiation Grant (Grant No. NTF23011). This research used resources of the Synergetic Extreme Condition User Facility
(SECUF, \url{https://cstr.cn/31123.02.SECUF}).
\end{acknowledgments}

\section*{Data Availability}
The data that support the findings of this study are available at
Zenodo~\cite{Yuan2026Data}. The code used to create the figures in
this paper is available from the authors on reasonable request.

\bibliography{references}
\end{document}